\documentclass[letter]{aa} 

\usepackage{multirow}
\usepackage{graphicx}
\usepackage{txfonts}

\usepackage[switch]{lineno}
 
\begin{document}

\newpage

   \title{Reshaping and ejection processes on rubble-pile asteroids from impacts}

   \author{S. D. Raducan\inst{1} \and
           M. Jutzi\inst{1} \and
           Y. Zhang\inst{2,4} \and
           J. Orm\"o\inst{3} \and
           P. Michel\inst{4}
          }

   \institute{Space Research and Planetary Sciences, Physikalisches Institut, University of Bern, Switzerland\\
              \email{sabina.raducan@unibe.ch}
         \and
             University of Maryland, College Park, USA
        \and 
            Centro de Astrobiologia (CAB), CSIC-INTA, Carretera de Ajalvir km 4, 28850 Torrejón de Ardoz, Madrid, Spain.
        \and 
            Université Côte d’Azur, Observatoire de la Côte d’Azur, CNRS, Laboratoire Lagrange, Nice, France
             }

   \date{Received August 25, 2022; accepted September 14, 2022}

 
  \abstract
   {Most small asteroids ($<$ 50 km in diameter) are the result of the breakup of a larger parent body and are often considered to be rubble-pile objects. Similar structures are expected for the secondaries of small asteroid binaries, including Dimorphos, the smaller component of the 65803 Didymos binary system and the target of NASA's Double Asteroid Redirection Test (DART) and ESA's Hera mission. The DART impact will occur on September 26, 2022, and will alter the orbital period of Dimorphos around Didymos. }
   {In this work we assume Dimorphos-like bodies with a rubble-pile structure and quantify the effects of boulder packing in its interior on the post-impact morphology, degree of shape change, and material ejection processes. }
   {We used the Bern smoothed particle hydrodynamics shock physics code to numerically model hypervelocity impacts on small, 160 m in diameter, rubble-pile asteroids with a variety of boulder distributions.}
   {We find that the post-impact target morphology is most sensitive to the mass fraction of boulders 
   comprising the target, while the asteroid deflection efficiency depends on both the mass fraction of boulders on the target and on the boulder size distribution close to the impact point. Our results may also have important implications for the structure of small asteroids.}
   {}


   \maketitle

\section{Introduction}

The structures and orbital properties of members of today's asteroid population in the Solar System have been altered by numerous collisional, dynamical, and thermal events. Most asteroids smaller than about 50 km in diameter are the result of the breakup of a larger parent body \citep{Bottke2005}. These objects are often `rubble piles'  \citep{Michel2001}, that is, aggregates held together only by self-gravity or small cohesive forces \citep{Richardson2002, Scheeres2010}. Although no direct measurement of an asteroid's internal structure has been performed yet, rubble-pile asteroids are expected to be composed of components ranging in size from several tens of metres down to millimetre particles \citep[e.g.][]{Fujiwara2006}. Recent images from the Hayabusa2 mission to Ryugu gave us the first glimpse into the structure of a small asteroid (870 m in diameter; \citealt{Arakawa2020, Sugita2019}), confirming predictions that bodies of this size may have a rubble-pile structure.

NASA's Double Asteroid Redirection Test (DART)  mission will impact Dimorphos, the secondary of the Didymos system, on September 26, 2022, and it will perform the first asteroid deflection test using a kinetic impactor. The deflection efficiency can be quantified in terms of a parameter called $\beta$, which is the ratio of the momentum of the impactor to the recoil momentum of the target. Past studies on the outcome of small-scale impacts on asteroid surfaces have shown that the impact outcome strongly depends on the surface, subsurface, and internal properties \citep[e.g.][]{Raducan2019, Raducan2020}. The impact ejecta plume created by DART will be imaged by the Italian CubeSat, LICIACube \citep{Dotto2021}. The mass, the surface morphology, and the interior structure of Didymos and Dimorphos will only be characterised four years later, by the Hera mission of the European Space Agency (ESA) \citep{Michel2020, Michel2022}; however, a rubble-pile structure is likely. 

In recent laboratory experiments, \cite{Ormo2022} conducted low velocity ($\approx$\,400 m/s) impacts on targets specifically designed to mimic rubble-pile asteroid surfaces composed of boulders and sand. The projectile was similar in size to boulders that were embedded into beach sand separated by the equivalent of one boulder diameter. They find that for their target set-up, the cratering displaces and ejects boulders rather than fragmenting them, unless they are directly hit. They also find that the ejecta curtains have higher ejection angles compared to homogeneous targets, and the boulders land at larger distances than the surrounding fine-grained material. Their results suggest that the ejection of boulders, as well as the magnitude and direction of the ejecta momentum vector, depends on the initial boulder distribution of the target. 

Here, we expand on the \cite{Ormo2022} impact experiments and numerically model impacts at real-asteroid scales on spherical, rubble-pile asteroids with different mass-frequency distributions of boulders. We aim to quantify the effects of boulder packing within the target on the post-impact morphology, the degree of shape change, and the ejection processes. The aim is to aid the analysis of the DART impact data, the future deflection of potentially hazardous asteroids, and the analysis of large craters on other rubble-pile asteroids. 


\section{Numerical methods}
We used the Bern parallel smoothed particle hydrodynamics (SPH) impact code \citep{Benz1995, Jutzi2008, Jutzi2015} to model DART-like impacts:  $\approx$\,600\,kg spherical aluminium projectiles impacting 160 m diameter spherical, rubble-pile targets at 6\,km/s. Recently, the code was validated against laboratory impact experiments on heterogeneous, rubble-pile-like targets \citep{Ormo2022}.

The composition and mechanical properties of asteroids vary from one body to another \citep{Tholen1984}. C-type asteroids (e.g. Ryugu and Bennu) have a composition similar to carbonaceous chondrites \citep{Yokoyama2022} and an average density of about 1.7 g/cm$^3$ (or even lower for rubble piles). The estimated tensile strength of the surface boulders was found to be considerably lower than measurements on meteorite samples \citep{Flynn2018}, for example 0.2--0.28 MPa on Ryugu \citep{Grott2019} and 0.44--1.7 MPa on Bennu \citep{Ballouz2020}. On the other hand, S-type asteroids (e.g. Itokawa and Didymos) have a composition similar to ordinary chondrites \citep{Nakamura2011} and a higher average density than C-types \citep{Carry2012, Pohl2020}. Moreover, for ordinary chondrites, the tensile strength seems to be about an order of magnitude higher than for carbonaceous chondrites \citep{Pohl2020}. However, despite the variation in mechanical properties of individual grains and boulders on different types of asteroids, the collective behaviour as a rubble pile could be similar, and the mechanisms studied here could be applicable for rubble-pile objects of various compositions. Therefore, without loss of generality, here we consider analogue materials that were recently used by \cite{Ormo2022} in their laboratory experiments. 

In our simulations, both the boulders and the granular target matrix were modelled using a Tillotson equation of state (EOS) for SiO$_2$ \citep[adapted from][]{Melosh2007}. The shear strength of the matrix material was described by the Drucker-Prager yield criterion (with an internal friction coefficient of $f$ = 0.56).
The boulders were modelled using a tensile strength and fracture model as described in \cite{Jutzi2015}, with parameters corresponding to a tensile strength of $Y_T$ $\approx$\,1 MPa. The initial porosity of the matrix and of the boulders was set to 35\% and was modelled using the $P-\alpha$ model with a simple power-law crush curve, informed by the experimental data for Lane Mtn. \#70 sand (\citealt{Housen2018}; Table~\ref{table:model_parameters}; Fig.~A.1). 
The projectile was modelled using the Tillotson EOS for aluminium and the von Mises yield criterion. The bulk density was $\rho$ = 1000 kg/m$^3$.

To compute the long-term evolution of the target (up to 2 hours after the impact), we used the `fast time integration scheme' described in \cite{Raducan2022c} (Appendix A.2). After the initial shock and fragmentation phase was over, we artificially changed the material properties of the target to a low sound speed medium, which allows larger time steps to be used. We reduced the bulk and shear moduli at three different times during the target evolution ($t_{transition}$ = 5, 50, and 500 s). Our SPH simulations had a spatial resolution of 2.5 million SPH particles in the target and 50 SPH particles in the projectile. 

\begin{table}[t!]
    \footnotesize
        \caption{Target material model parameters for our impact simulations.}
        \begin{tabular}{lll}
    \bf{Description}        & \bf{Boulders} & \bf{Target matrix}  \\
    \hline
    Material             & SiO$_2$ & SiO$_2$ \\
        Equation of state (EOS)    & Tillotson$^{a,b}$ & Tillotson$^{a,b}$ \\
        Yield criterion             & Tensile$^c$ & Drucker-Prager \\
        Initial bulk modulus, A (GPa) & 35.9 & 35.9 \\   
        \hline
        \bf{Drucker-Prager criterion}\\
        Cohesion, $Y$ (Pa)   & 10$^7$ & 0 \\
        Strength at infinite pressure, $Y_{m}$ (GPa)  & 1.0 & 0.1  \\
        Internal friction coefficient, $f$    & 0.8 & 0.56 \\
        \hline
        \bf{Porosity model ($P-\alpha$)$^d$}           \\ 
        Initial distension, $\alpha_0$          & 1.55 & 1.55  \\
        Solid pressure, $P_s$ (MPa)                             & 40 & 20  \\
        Elastic pressure, $P_e$ (MPa)                             & 1 & 1 \\
        Exponent, $n$                                   & 2 & 2 \\
        \hline
    \multicolumn{3}{l}{
    $^a$\cite{Tillotson1962};
    $^b$\cite{Melosh2007};
    $^c$\cite{Jutzi2015};
    $^d$\cite{Jutzi2008}.}
        \end{tabular}
        \label{table:model_parameters}
\end{table}

In order to obtain realistic configurations of boulders, we used the $N$-body tree code pkdgrav \citep{Richardson2000} and its soft-sphere discrete element method (SSDEM) framework \citep{Schwartz2012, Zhang2017}. We simulated the gravitational collapse of a cloud of spherical particles with a predefined size distribution, and then we shaved extra particles off the accumulated aggregate to represent a 160 m spherical body. To explore a large possible range of boulder mass fractions, we removed some of these boulders from pkdgrav output when we built our SPH models. The voids between boulders were filled with matrix material. 

We defined two categories of targets, depending on the ratio of boulder material mass to matrix material mass (packing): The first is high packing ($>$35\% boulder mass fraction). We considered targets made from boulders with radii of: (1)  7.5 m  (52.36\% boulder mass fraction); (2) 2.5 to 14 m, with a size-frequency distribution following a differential power law with an exponent of $-3$, similar to that on the asteroid Itokawa (38.18\% boulder mass fraction); and (3) 2.5 m (35.7\% boulder mass fraction).

The second category is loose packing ($<$20\% boulder mass fraction). We considered targets made from (4) 2.5 to 12 m boulders with loose packing (17.5\% boulder mass fraction); and (5) 5 m boulders (16.17\% boulder mass fraction). 

For each of these targets, we chose four different impact locations. The different simulations are marked as A, B, C, and D  in Fig.~\ref{fig:final} (see Appendix D).
The minimum boulder size of 2.5 m in radius is the lower limit for which we can properly resolve the boulders ($\approx$\,30 SPH particles in each boulder for a resolution of 2.5 million SPH particles in the target).

\vspace{-0.2cm}
\section{Results}

\begin{figure*}[h!]
        \centering
        \includegraphics[width=0.74\linewidth]{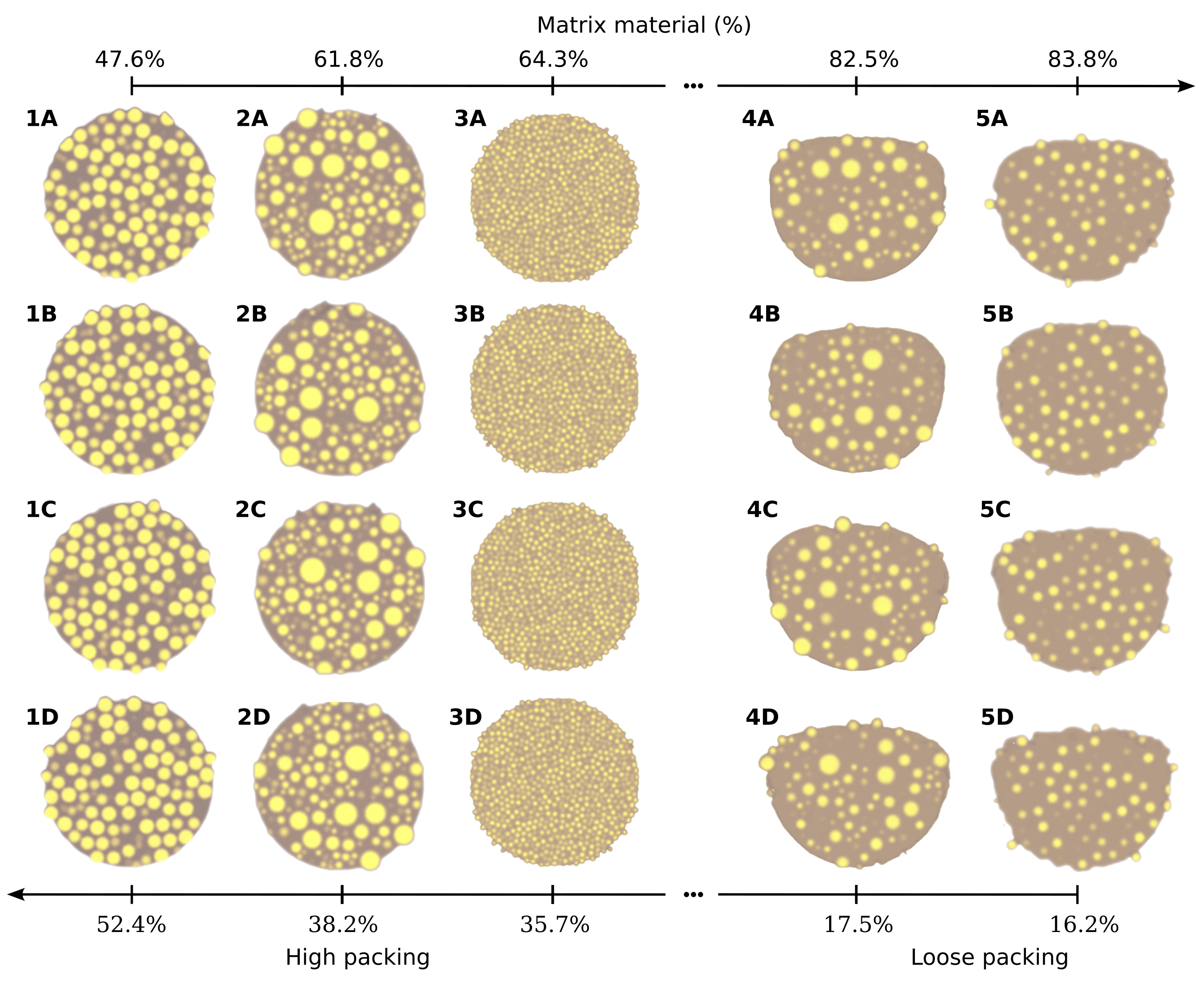}
        \vspace{-0.2cm}
        \caption{Two-dimensional slices showing possible asteroid morphologies after impacts on initially 160 m spherical bodies with varying boulder size-frequency distributions. The impacts on targets with a high packing of boulders produce impact craters, while impacts on targets with a loose packing of boulders produce a global deformation of the target. Due to the resolution employed in our simulations (see Sect. 2), the size and morphology of the impact craters was not determined.} 
        \label{fig:final}
\end{figure*}

\subsection{Target morphology}

Figure~\ref{fig:final} shows two-dimensional slices through the post-impact asteroid morphology resulting from our impact simulations on targets with different boulder size-frequency distributions. In the high packing cases (i.e. when more than 35\% of the target mass fraction is occupied by boulders; Fig.~\ref{fig:final}, high packing), the impact creates a very small crater, only a few times larger than the projectile. These craters are under-resolved in our simulations. On the other hand, in target scenarios with a loose packing of boulders, the impact causes a global deformation of the target (Fig.~\ref{fig:final}, loose packing). Similar target deformations are seen from impacts on homogeneous targets made of solely matrix material (\cite{Raducan2022c}; Appendix B).

We find that the impact location has little influence on the post-impact morphology of the target. For example, in the case 5B, the projectile struck a 5 m boulder, and in the case of 5C it hit at the edge of a 5 m boulder, while in cases 5A and 5D there were no boulders close to the impact site. However, the resulting target morphologies from these four impact scenarios are very similar (Fig.~\ref{fig:final}). Our results suggest that the post-impact asteroid morphology is dictated by the number density of boulders and the associated interlocking effect, which -- depending on the packing -- can hinder the crater growth.

\subsection{Mass-velocity distribution of the ejecta}

\begin{figure*}[h!]
        \centering
        \includegraphics[width=\linewidth]{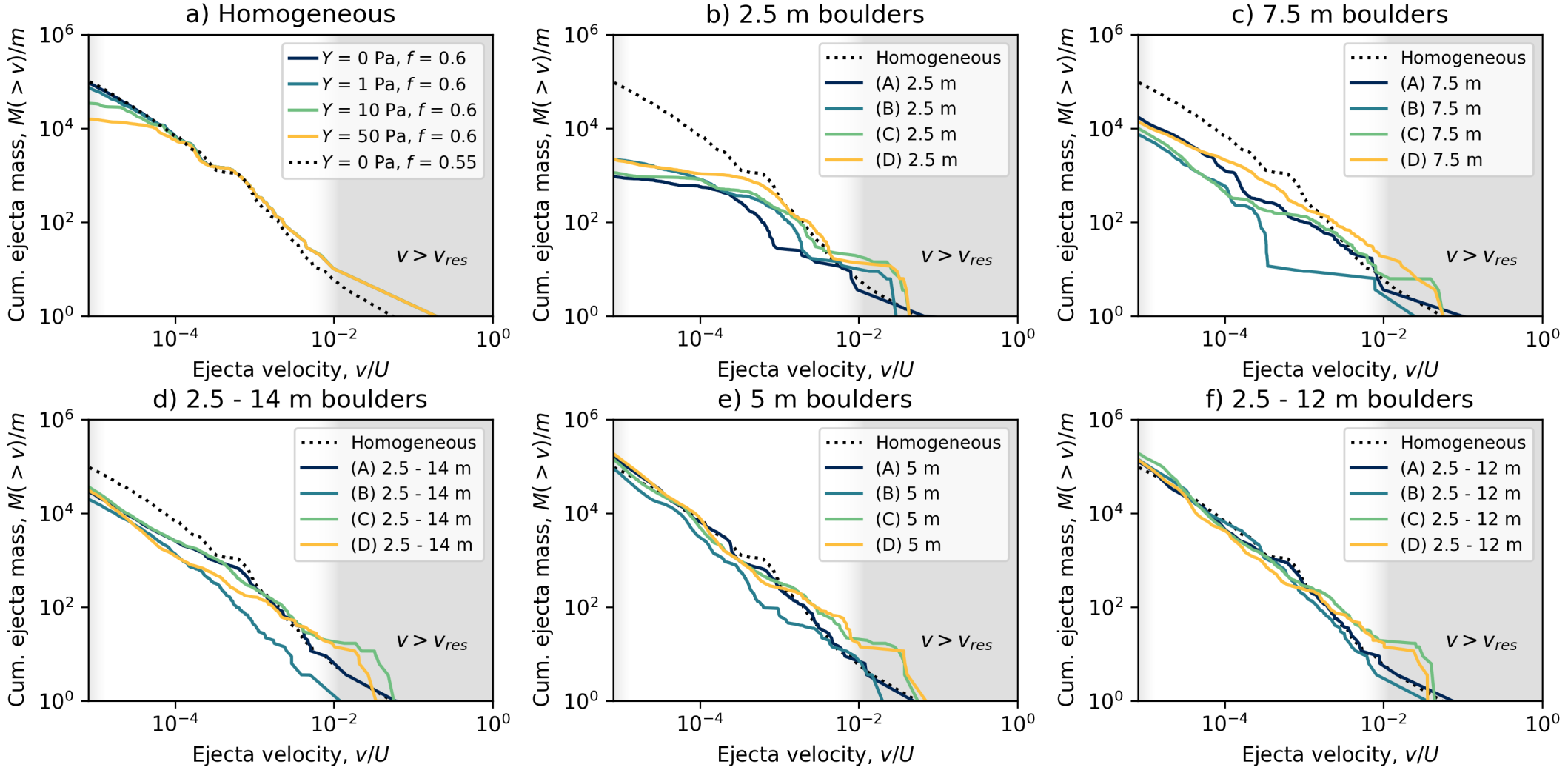}
        \caption{Cumulative ejecta mass normalised by the projectile mass, $m$, ejected at speeds greater than $v$ as a function of normalised ejection speed, $v/U$, for impacts on various targets.\ Panel (a): Homogeneous targets with $Y$ = 0 Pa, $f$ = 0.55 and results from \cite{Raducan2022c} of impacts on homogeneous targets with varying cohesion ($Y$ = 0 -- 50 Pa, $f$ = 0.6). Panels (b), (c), (d), (e), and  (f): Targets with different boulder configurations. Panels (b), (c), and (d) are targets with high boulder packing and (e) and (f) are targets with loose boulder packing. In all panels, the dashed line shows the mass-velocity distribution from a 600 kg impact at 6 km/s on a homogeneous (matrix-only) target. The shaded area shows the velocity range for which the ejecta might be under-resolved (see Sect. 2).} 
        \label{fig:m_v}
\end{figure*}

Figure~\ref{fig:m_v} shows the normalised cumulative ejecta mass (where $m$ is the projectile mass), as a function of normalised ejecta velocity (where $U$ is the impact velocity), from impact simulations with different boulder configurations. In Fig.~\ref{fig:m_v}a we plot the mass-velocity distribution from an impact on a cohesionless, homogeneous target (made of only matrix material, $Y$ = 0 Pa, $f$ = 0.55). The results are in good agreement with mass-velocity distributions from the impacts on homogeneous spherical targets studied in \cite{Raducan2022c} ($Y$ = 0 -- 50 Pa, and $f$ = 0.6). We used the results from the homogeneous target to compare with our results from impacts on rubble-pile targets with different configurations of boulders (Fig.~\ref{fig:m_v}b, c, d, e, and f). In the heterogeneous target scenarios, the ejected mass contains both boulder and matrix material. 

The amount of total ejected material from the impacts studied here is most sensitive to the boulder packing and to the size of the boulders within the target, and less sensitive to the impact location. In the target scenarios with 2.5 m boulders (Fig~\ref{fig:m_v}b), the total mass of ejecta, $\sum M/m$ (where $M$ is the mass of ejecta above escape speed and $m$ is the projectile mass), is between 50 and 100 times lower than in a homogeneous target scenario. For impacts on targets with 7.5 m boulders (Fig~\ref{fig:m_v}c) and targets with 2.5 to 12 m boulders (Fig~\ref{fig:m_v}d), $\sum M/m$ is about ten times lower. For the targets with loose boulder packing (Fig.~\ref{fig:m_v}e, f), $\sum M/m$ is of the same order of magnitude as the total ejected mass from a homogeneous target scenario. In all cases, the impact location causes a spread in $\sum M/m$ of up to 40\%.

\subsection{Momentum enhancement}

\begin{figure}[h!]
        \centering
        \includegraphics[width=\linewidth]{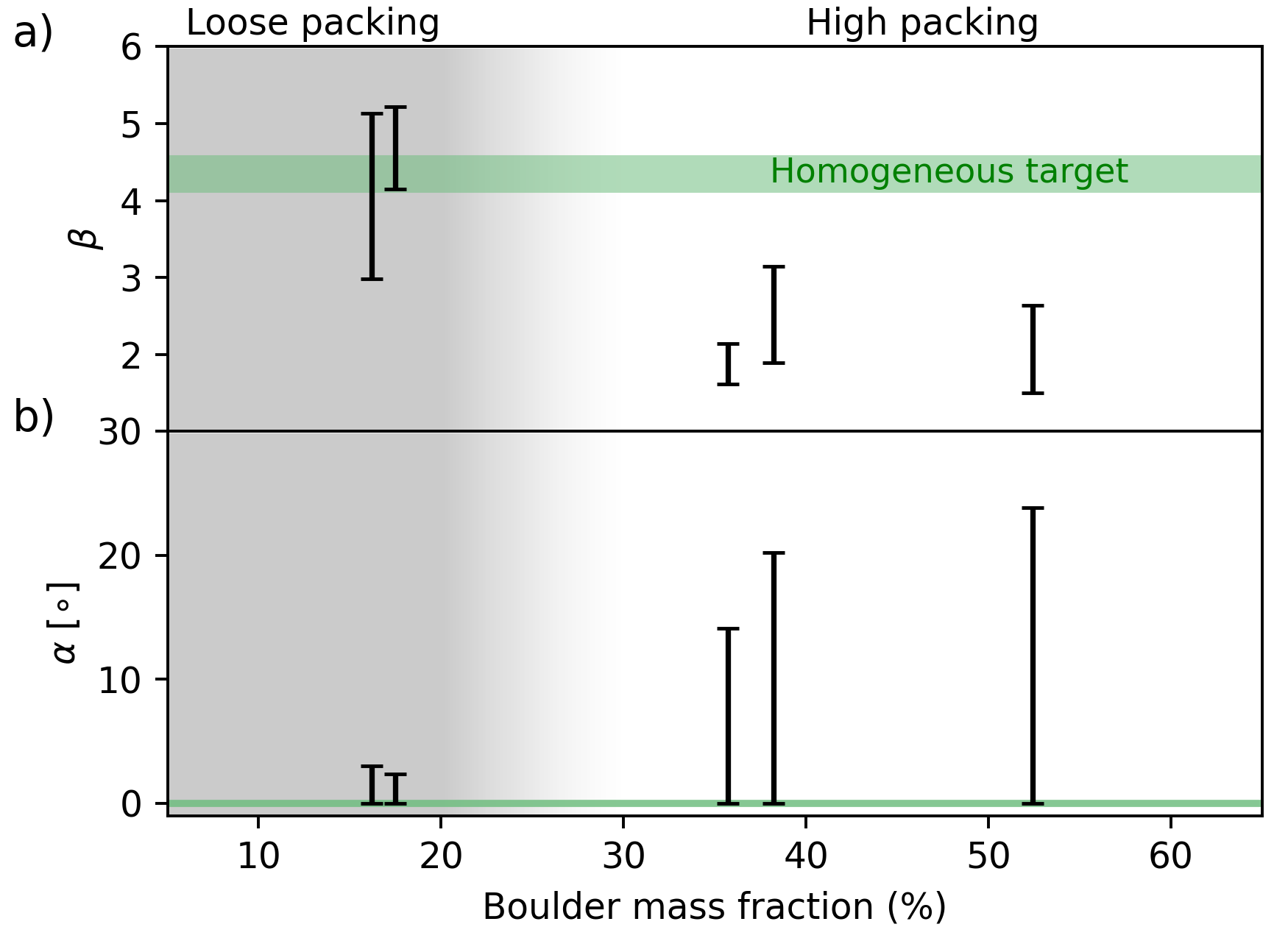}
        \caption{Momentum transfer efficiency and the direction of the direction of the ejecta momentum vector. Panel (a): Momentum transfer efficiency, $\beta$, for different boulder mass fractions. Panel (b): Angle between the surface normal and the ejecta momentum vector, $\alpha$. The shaded area denotes the loose packing regime. The horizontal green band shows the momentum enhancement from equivalent impacts on a homogeneous, matrix-only target.} 
        \label{fig:beta}
\end{figure}

The ejecta mass-velocity distributions shown in Fig.~\ref{fig:m_v} were integrated to determine the cumulative, vertically ejected momentum, $p_{ej(z)}/mU = \beta-1$, where $m$ and $U$ are the mass and the velocity of the projectile, respectively (Appendix C; Fig.~\ref{fig:beta}a). For the high packing target scenarios, we find $\beta$ to be between $\approx$\,1.5 and $\approx$\,3.2. These values are up to 60\% smaller than the momentum enhancement resulting from an equivalent impact on a homogeneous target ($\beta$ = 4.3). On the other hand, in the loose packing target scenarios, the presence of boulders near the impact point can cause both a reduction (up to 35\%) and amplification (up to 15\%) in $\beta$. For each target boulder configuration studied here, the spread in $\beta$ due to the impact location was about 60\%. This suggests that both the boulder mass fraction within the target and the impact location (or whether the projectile hit a boulder or not) are important when quantifying the momentum enhancement factor, $\beta$. 

For vertical impacts on homogeneous targets, the impact ejecta is symmetric, and the momentum enhancement vector is parallel to the surface normal \citep[e.g.][]{Rivkin2021}. We find that the presence of the boulders causes a deviation in the momentum enhancement vector from the normal of up to $\alpha\approx$\,25$^\circ$ (Fig.~\ref{fig:beta}b). However, $\alpha$ decreases with decreasing boulder mass fraction within the target. For the targets with loose packing studied here, $\alpha<$3$^\circ$.

\subsection{Ejected boulders}

\begin{figure*}[t!]
        \centering
        \includegraphics[width=0.94\linewidth]{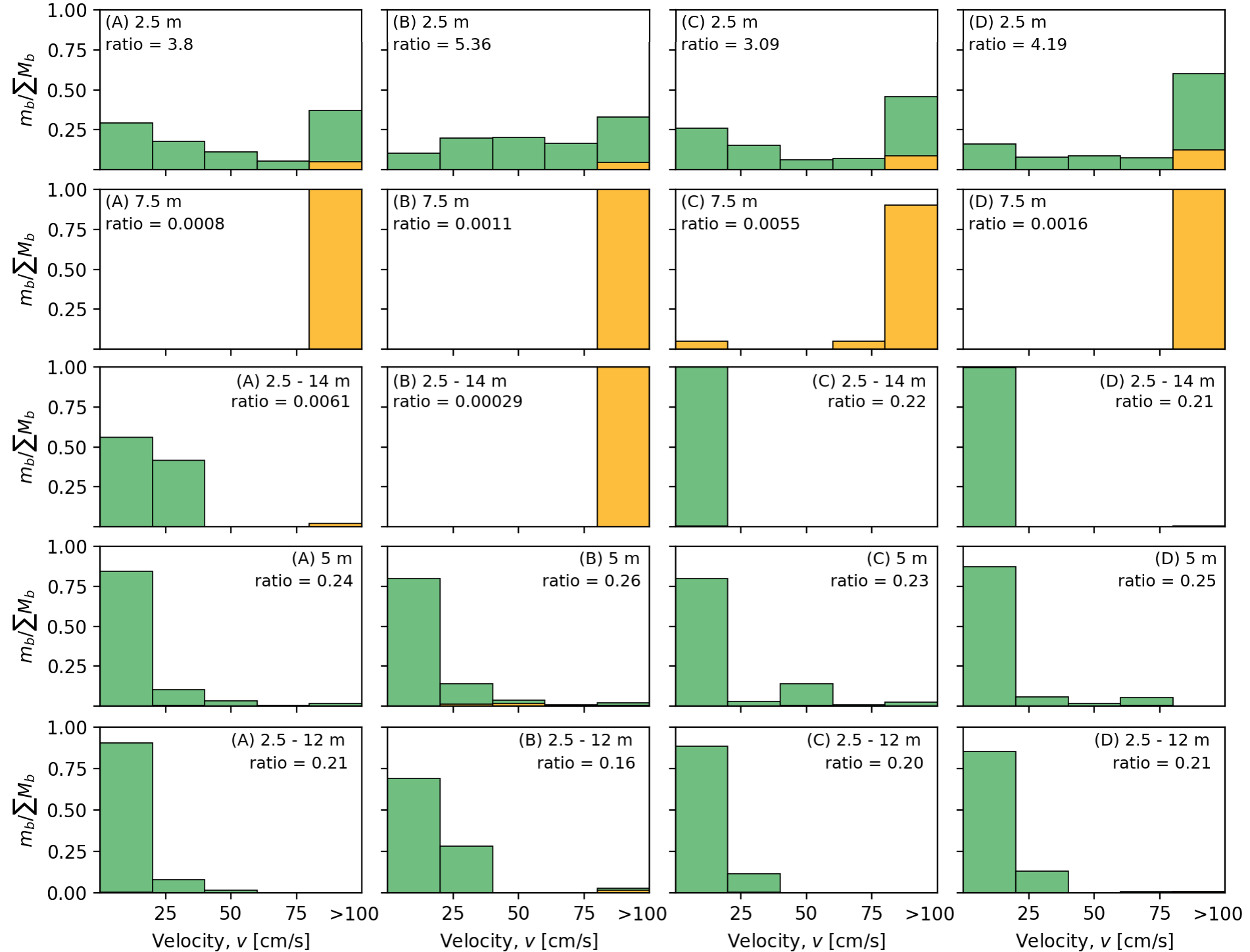}
        \caption{Histogram of the mass fraction of boulders ejected with speeds above the target escape velocity ($v>v_{esc}$), for targets with different configurations of boulders. $m_b$ is the boulder mass in each bin, and $\sum M_b$ is the total mass of ejected boulder material above $v>v_{esc}$. The velocity data are discretised into 20 cm/s bins. The total mass of boulder material ejected is shown in green, and the mass of ejected boulder material that has been fragmented ($D>$ 0.5) is shown in yellow. In each panel we show the ratio of the sum of ejected boulder mass to the sum of total ejecta mass (ratio = $\sum M_b/\sum M$).}
        \label{fig:hist}
\end{figure*}

In our numerical simulations of impacts on rubble-pile targets, most of the ejected boulders are intact (damage $D$ = 0) and only the boulders initially close to the impact point are fragmented ($D >$ 0.5; \citealt[e.g.][]{Jutzi2015}). 
Figure~\ref{fig:hist} shows the mass fraction of boulder material ejected at speeds higher than the target escape velocity ($v>v_{esc}$). Our results show that high velocity impacts on rubble-pile targets with 2.5 m boulders eject mostly intact boulders, and only between 5 and 15\% of the ejected boulder material is damaged. While the intact boulder material is ejected at a range of velocities (from $v_{esc}$ to more than 100 cm/s), the fragmented material is only ejected at relatively high velocities ($v>$ 100 cm/s), compared to $v_{esc} \approx$\,5 cm/s. In this impact scenario, the boulders are only $\approx$\,5 times larger than the projectile. The fragmented boulder material is ejected at higher velocities compared to the intact boulders because the fragmented material originates close to the impact point, where the pressure gradient is larger than it is farther away from the impact. On the other hand, in the case of impacts on rubble-pile targets with 7.5 m boulders, the boulders are about 14 times larger than the projectile and therefore too massive to be ejected intact. In these impacts, all the ejected boulder material is damaged and most of it leaves the target at speeds above 100 cm/s. In the impact simulations on rubble-pile targets with 2.5 to 14 m boulders, the ejection of boulders depends on the size and location of the boulders around the impact point. The ejection of boulders in the case of the rubble-pile targets with loose packing (5 m boulders and 2.5 - 12 m boulders) is less sensitive to the impact location. In these impact models, fewer than 2\% of boulders are fragmented and most of the intact boulders are ejected at low speeds ($v <$ 40 cm/s). Overall, our results imply that boulders larger than $\approx$\,10 m in diameter are ejected with speeds lower than 100 cm/s.

\vspace{-0.2cm}
\section{Discussion}

Our simulation results of DART-like impacts on small rubble-pile asteroids show that the outcome of such impacts is sensitive to both the local and the global distribution of boulders within the body. Moreover, the post-impact target morphology is a diagnosis of the interior boulder distribution. In the case of the DART impact on Dimorphos -- assuming that the asteroid has similar dimensions and masses as the targets modelled here -- if the impact creates a crater on the surface, then the boulder packing is more than 35\% and/or the cohesion is larger than that in our study \citep{Raducan2022c}. On the other hand, if the DART impact causes the global deformation of the asteroid, then the interior of Dimorphos has a low cohesion and a loose packing of boulders. 
The ejecta plume images expected to be obtained by LICIACube \citep{Dotto2021} might also give an insight into the boulder packing within the target. As shown by our simulation results, in impacts on targets with a high packing of boulders, there is less ejected mass than in impacts involving targets with a loose packing of boulders. Moreover, for a high packing of boulders, the crater growth ceases after just a few tens of seconds. Therefore, if the ejecta plume is still opaque at $\approx$\,160 s after the impact and is mostly homogeneous, then the target is most likely weak ($Y_0 \lessapprox$ 1 kPa; e.g. \citealt{Cheng2022}) and with a loose packing of boulders. 

One of the main measurable quantities from the DART impact is the change in orbital velocity of Dimorphos, which is strongly related to the magnitude and direction of the ejecta momentum vector \citep{Rivkin2021}. We show that $\beta$ is sensitive to both the boulder packing and to the size-frequency distribution of boulders at the impact site. Though images of the target may be available before the impact, our results provide further insight into its properties and configuration. If large values of $\beta$ are estimated, then we can predict that the target is most likely weak, with a loose boulder packing, and that the impact did not hit a large boulder (i.e. the energy required to disrupt the boulder is much smaller than the impact energy). 

In our impact simulations, boulders can be ejected from the target, mostly intact. We find that a mass of $\approx$\,10$^4 \times$ $m$ (where $m$ is the projectile mass) is ejected as boulders larger than 5 m in diameter (e.g. Figs.~\ref{fig:m_v} and \ref{fig:hist}). In the case of the DART impact, such boulders may leave the asteroid system, orbit the system, or land on Didymos or back on Dimorphos \citep{Yu2018}. The ejection of large intact boulders in the impact regime investigated here may also have important implications for the structure of small asteroids.
Our results imply that sub-catastrophic impacts on larger asteroids can eject large intact boulders, which could become monolithic asteroids.

Our results also show that the boulder mass-fraction on an asteroid can evolve over time, depending on its collisional history. For example, in the target scenario with 2.5 m boulders, the impact ejects preferentially more boulder material than matrix material (Fig.~\ref{fig:hist}), while on other targets the impact ejects more matrix material than boulder material. This mechanism may explain why some asteroids are more boulder rich (e.g. Ryugu and Bennu) than others (e.g. Itokawa and Eros).

\vspace{-0.2cm}
\section{Conclusions}

We studied DART-scale impacts on rubble-pile asteroids with different boulder size-frequency distributions. We show that the outcome of the impact can be a diagnosis of the asteroid's internal structure. If the impact results in a global deformation of the target, the ejecta plume is still present at $\approx$\,160 s after the impact, and a large momentum enhancement resulting from the impact (i.e. a large $\beta$) is estimated, then Dimorphos is most likely very weak and with a loose packing of boulders. On the other hand, if the impact creates a very small crater (only a few times larger than the spacecraft), there is no observed ejecta plume, and $\beta$ is small, then it does not necessarily imply that the target is strong, as currently believed. Instead, it is still possible that the target has a low-cohesion matrix but a high boulder packing. Our modelling results together with future measurements by the Hera mission will provide constraints regarding the evolution of small rubble-pile asteroids by sub-catastrophic impacts.
   
\begin{acknowledgements}
      This work has received funding from the European Union’s Horizon 2020 research and innovation programme under grant agreement No. 870377. P.M. also thanks ESA, CNES and the CNRS through the MITI interdisciplinary programs for funding support.JO was supported by the Spanish State Research Agency (AEI) Project No. MDM-2017-0737 Unidad de Excelencia “María de Maeztu” – Centro de Astrobiología (CSIC-INTA).
\end{acknowledgements}

%
%

\bibliography{refs.bib} 
\bibliographystyle{aa}  

\appendix

\section{Numerical method}

In this work we used the Bern SPH shock physics code \citep{Benz1995, Jutzi2008, Jutzi2015}. The code was originally developed by \cite{Benz1995, Benz1999} to model the collisional fragmentation of rocky bodies. It was later parallelised \citep{Nyffeler2004} and further extended by \cite{Jutzi2008}, \cite{Jutzi2014}, and \cite{Jutzi2015} to model porous and granular materials. The most recent version of the code includes a tensile fracture model \citep{Benz1995}, a porosity model based on the P-$\alpha$ model \citep{Jutzi2008, Jutzi2009}, pressure-dependent strength models \citep{Jutzi2015}, and self-gravity. 

\subsection{Porosity model}

The initial porosity of the boulders and of the target matrix was set to 35\% and was modelled using the $P-\alpha$ porosity compaction model \citep{Jutzi2008}. The full description of the $P-\alpha$ model implemented in the Bern SPH code is given by \cite{Jutzi2008}. 

Here we used a simplified version of the $P-\alpha$ model, with a single power-law slope, defined by the solid pressure, $P_s$, elastic pressure, $P_e$, exponent, $n$, initial distension, $\alpha_0$, and distension at the transition from the elastic regime, $\alpha_e$:

\begin{equation}
  \alpha(P)=\begin{cases}
    1, & \text{if $P_s<P$}\\
    (\alpha_e-1)\left(\frac{P_s-P}{P_s-P_e}\right)^{n}+1, & \text{otherwise.}
  \end{cases}
\end{equation}

\begin{figure}[h!]
        \centering
        \includegraphics[width=\linewidth]{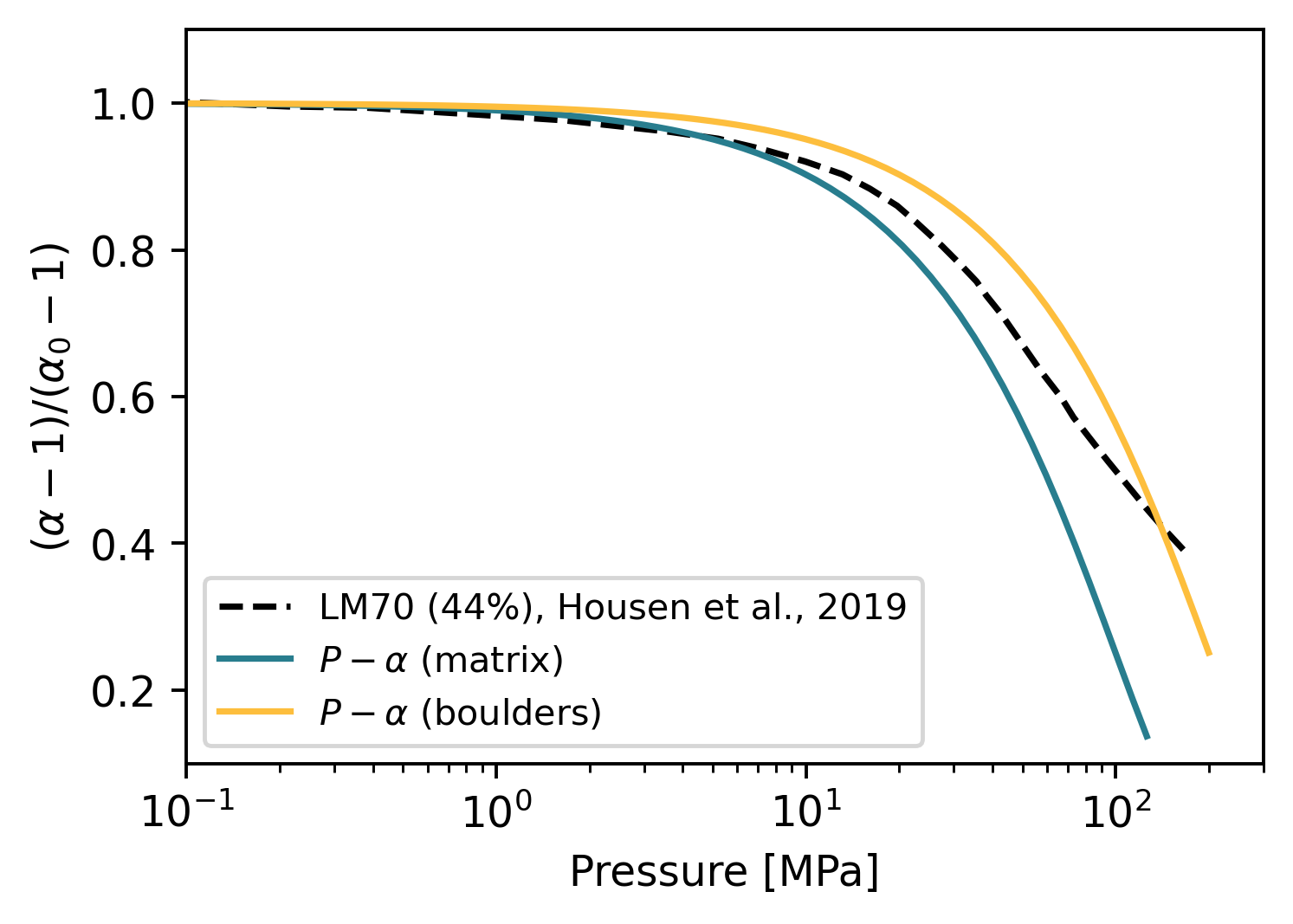}
        \caption{Normalised crush curves for Lane Mtn. \#70 sand \citep{Housen2018} and for the matrix and boulder material used in our numerical simulations.}
        \label{fig:porosity}
\end{figure}

The input parameters from the sand matrix and for the boulders are summarised in Table~\ref{table:model_parameters}. We assume that $\alpha_e = \alpha_0$. The input parameters were informed by the experimental data crush curve for Lane Mtn. \#70 sand \citep{Housen2018}, which had a porosity of 44\%. In Fig.~\ref{fig:porosity} we compare the crush curve for Lane Mtn. \#70 sand with our $P-\alpha$ crush curves for the matrix and boulders. The figure shows the normalised crush curves such that the value is 1 for the initial state and 0 when the material is crushed to its solid density (see \citealt{Housen2018} for details). 

\subsection{Modelling approach for the late-stage evolution}
This study was motivated by the DART impact on Dimorphos, and therefore we simulated DART-like impacts on small, Dimorphos-like targets \citep{Naidu2020}. Due to the vastly different timescales of the shock-wave propagation and the crater formation, it is difficult to numerically model the entire crater formation on small ($\sim$ 100\,m -- 1000\,m), weak asteroids. 
To ensure numerical stability, the maximum time step in a shock physics code is limited by the Courant criterion. In SPH, the Courant criterion requires that the time step, $dt$, is smaller than the simulation resolution divided by the sound speed in the target, $c_s$. To model DART-like impacts on Dimorphos, the maximum time step must therefore be $dt < \mathrm{resolution} /c_s \simeq 10^{-4}$\,s. On the other hand, the crater formation and ejecta re-accumulation time in the gravity regime can last up to a few hours. 

Here we applied in the shock physics code calculation a transition to a low-speed medium. 
At a time $t_{transition}$ after the initial shock has passed, the target material movement is governed only by low velocities, $v_{material}<<c_s$, and we can switch to a low-speed medium (i.e. a fast time integration scheme). In this step, we applied a simplified Tillotson EOS, in which all energy related terms are set to zero. The remaining leading term of the EOS is governed by the bulk modulus $P=A(\rho/\rho_0-1)$, which also determines the magnitude of the sound speed. At $t_{transition}$ = 5, 50, and 500 s, we used $A$ = 359\,MPa, $A$ = 3.59\,MPa, and $A$ = 35.9\,kPa, respectively. The shear modulus was also reduced proportionally. This approach has been tested extensively in parameter studies (unpublished), benchmarked against other shock physics codes (i.e. \citealt{Luther2022}), and validated against laboratory experiments with homogeneous and heterogeneous targets (e.g. \citealt{Ormo2022, Raducan2022c}).

\section{Homogeneous target results}

\begin{figure}[h!]
        \centering
        \includegraphics[width=\linewidth]{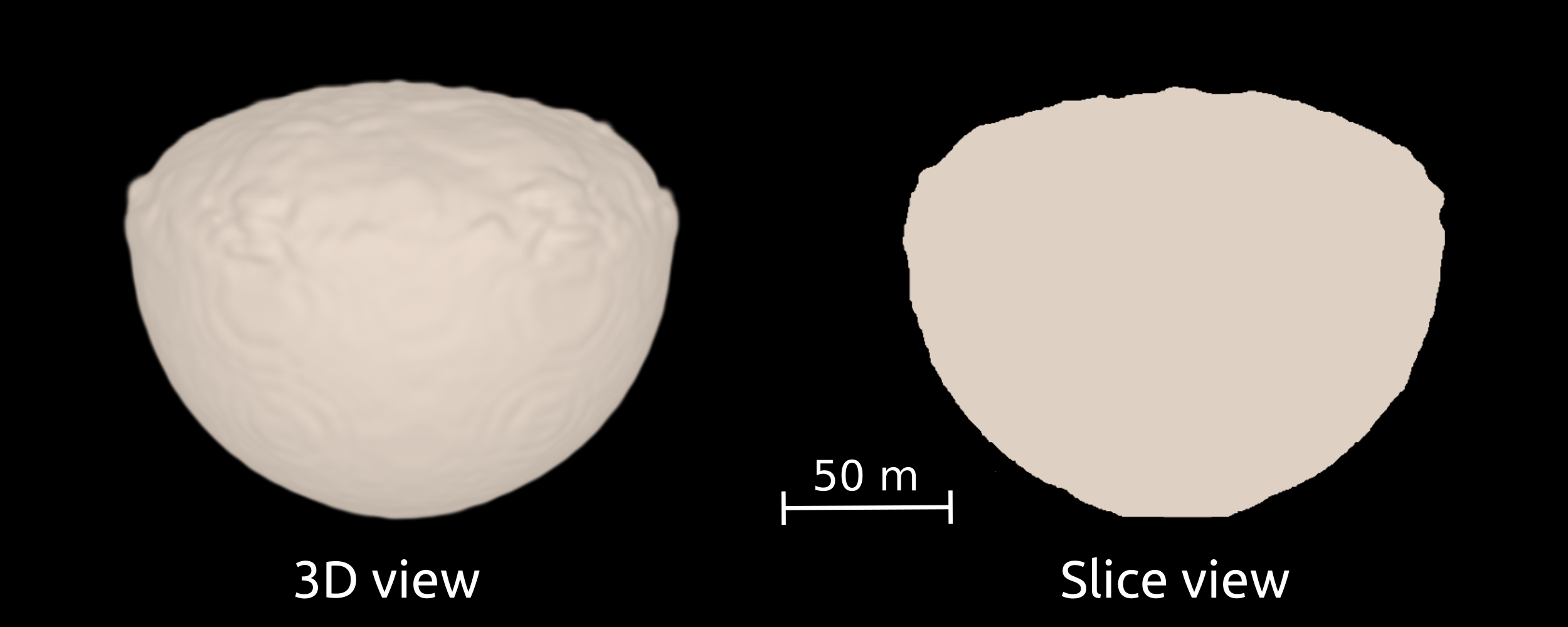}
        \caption{Three-dimensional and two-dimensional slice through the post-impact target morphology of an initially homogeneous, spherical asteroid (160 m in diameter).} 
        \label{fig:hh}
\end{figure}

In this work we compare the outcome of DART-like impacts on rubble-pile targets with the outcome of an equivalent impact on a homogeneous, matrix-only spherical target (160 m in diameter). Figure~\ref{fig:hh} shows a three-dimensional view and a two-dimensional slice through the post-impact morphology of an initially homogeneous, spherical target. The impact causes the global deformation of the asteroid. These types of sub-catastrophic impacts have been studied in detail in \cite{Raducan2022c}.

\begin{figure*}[h!]
        \centering
        \includegraphics[width=\linewidth]{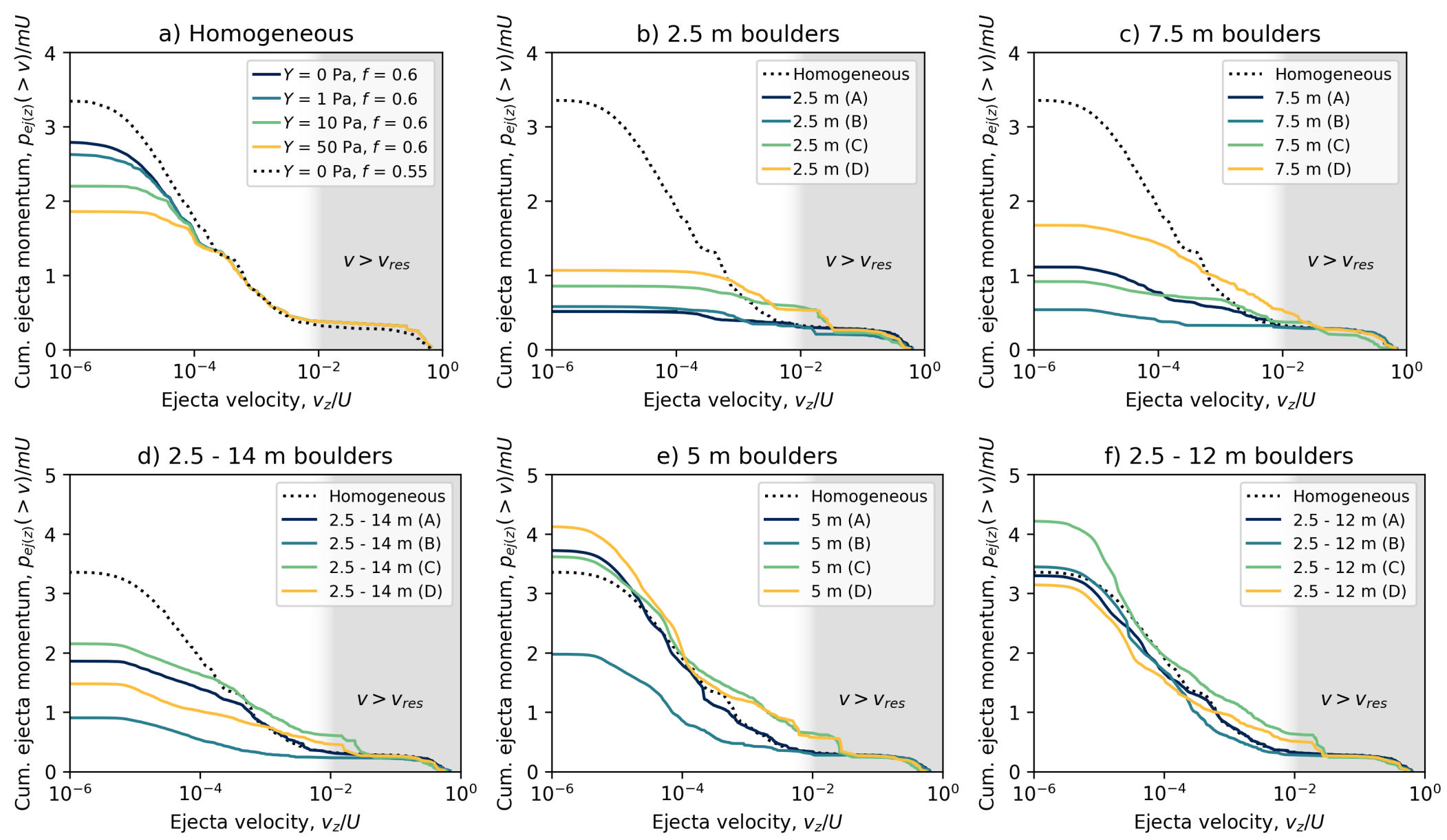}
        \caption{Cumulative ejecta momentum as a function of normalised vertical ejection velocity for impacts on various targets.\ Panel (a): Homogeneous targets with varying cohesion and coefficient of internal friction (adapted from \citealt{Raducan2022c}) and targets with different boulder configurations. Panels (b), (c), and (d): Targets with high boulder packing. Panels (e) and (f): Targets with loose boulder packing. In all panels, the dashed line shows the cumulative ejecta momentum from an impact on a homogeneous (matrix-only) target.} 
        \label{fig:p_v}
\end{figure*}

\section{Momentum enhancement calculations}\label{ch:beta}

In a high velocity impact event, the change in momentum of the asteroid, $\Delta \boldsymbol{P}$, is amplified by the momentum of impact ejecta that escapes the gravitational attraction of the target body. In a vertical impact, the change in linear momentum of the asteroid as a result of the impact, $\Delta\boldsymbol{P}$, is therefore equal to the impactor momentum, $m\boldsymbol{U}$, plus the resultant momentum of the escaping ejecta, $\boldsymbol{p_{ej}}$. The total momentum change of the asteroid in the direction perpendicular to the asteroid’s surface at the point of impact divided by the impactor momentum is a measure of deflection efficiency, commonly defined as $\beta = \Delta P_z/(m U)$ \citep{Housen2012}. The full equation needed to determine the momentum multiplication factor, $\beta$, from observations is derived in \citet{Rivkin2021}. For a vertical impact, the equation can be simplified to
\begin{equation}
    \beta = \frac{\Delta {P_z}}{mU} = 1 + \frac{{p_{ej(z)}}}{m U}.
\end{equation}
The value of $\beta$ depends on both the target material properties and the impact conditions \citep[e.g.][]{Jutzi2014, Syal2016, Stickle2015, Raducan2019, Raducan2021}. A value of $\beta\approx$\,1 implies that the crater ejecta makes a negligible contribution to the total momentum change, while $\beta>2$ means that the momentum contribution from the crater ejecta is larger than the momentum imparted by the impactor directly.

Here $\beta$ is calculated by summing over all the SPH particles with ejection velocities larger than $v_{esc}$. For a given vertical impact, the ejecta momentum is given by 
\begin{equation}
    {p_{ej(z)}} = \sum m_e \times v_z,
\end{equation}
where $m_e$ and $v_z$ are the mass and vertical velocity of individual SPH particles, respectively. In the impact simulations presented here, we were able to track the impact ejecta and to perform the $\beta$ calculation after long periods (up to 2\,h after the impact). Therefore, the ${p_{ej(z)}}$ calculation takes the gravitational influence of the target into account. 

The ejecta mass-velocity distributions in Fig.~\ref{fig:m_v} were integrated to determine the cumulative, vertically ejected momentum, $p_{ej(z)}/mU = (\beta-1)$. Figure~\ref{fig:p_v} shows the cumulative ejecta momentum normalised by the projectile momentum ($mU$) as a function of vertical ejection velocity, $v_z/U$, where $U$ is the impact velocity.

\section{Heterogeneous target input geometries}

\begin{figure*}[t!]
        \centering
        \includegraphics[width=\linewidth]{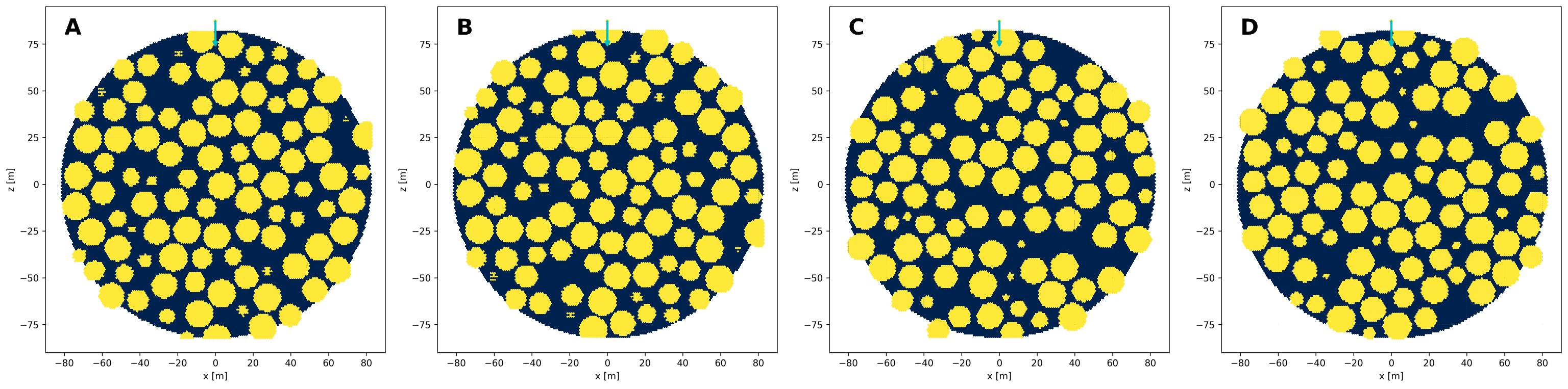}
        \caption{Slices through the initial target geometries with 7.5 m boulders (high boulder packing). The arrow at the top of each panel shows the impact location. } 
        \label{fig:init_800}
\end{figure*}

\begin{figure*}[h!]
        \centering
        \includegraphics[width=\linewidth]{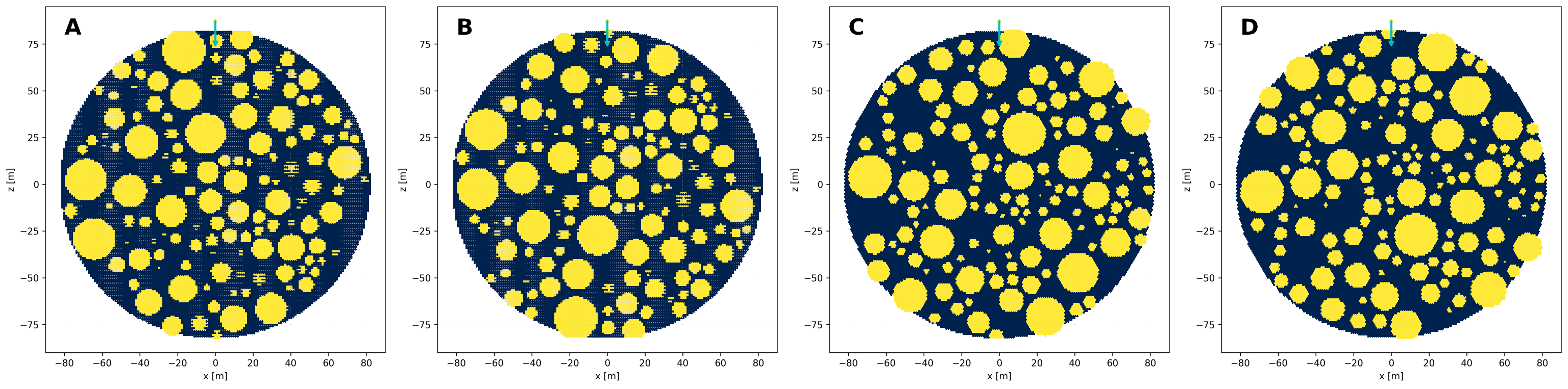}
        \caption{Slices through the initial target geometries with 2.5 to 12 m boulders (high boulder packing). The arrow at the top of each panel shows the impact location. } 
        \label{fig:init_ryu}
\end{figure*}

\begin{figure*}[h!]
        \centering
        \includegraphics[width=\linewidth]{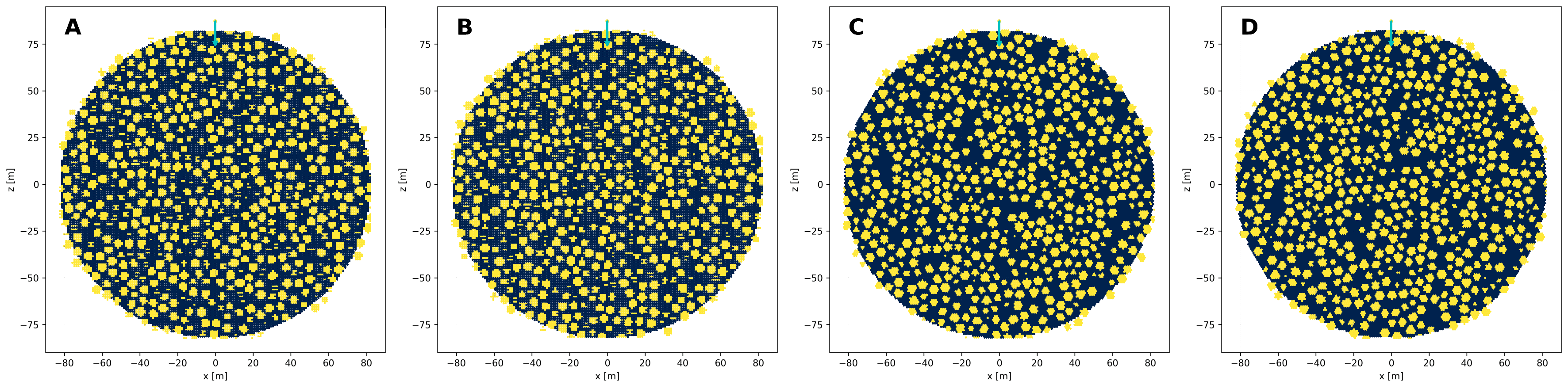}
        \caption{Slices through the initial target geometries with 2.5 m boulders (high boulder packing). The arrow at the top of each panel shows the impact location. } 
        \label{fig:init_300}
\end{figure*}

\begin{figure*}[t!]
        \centering
        \includegraphics[width=\linewidth]{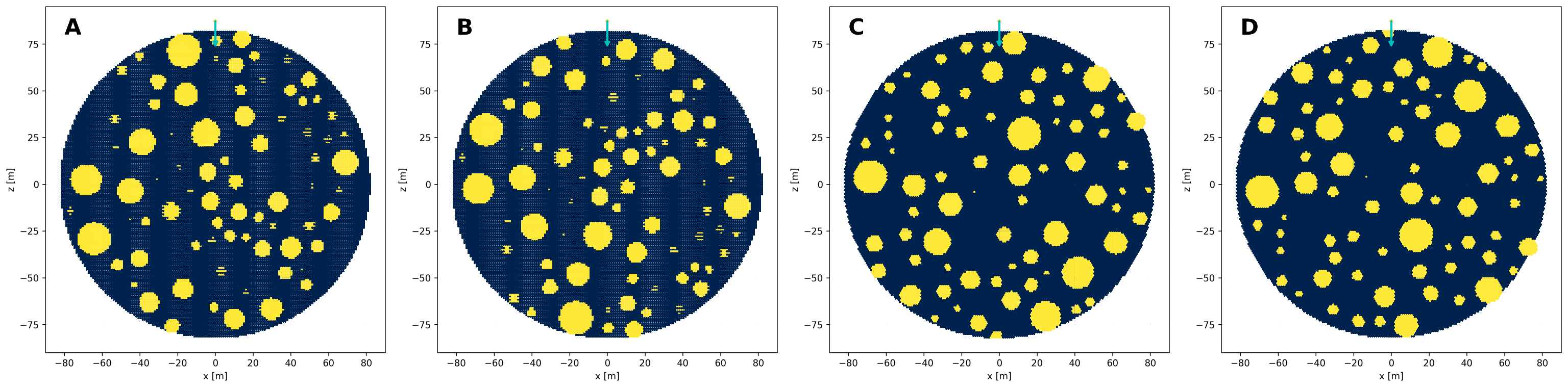}
        
        \includegraphics[width=\linewidth]{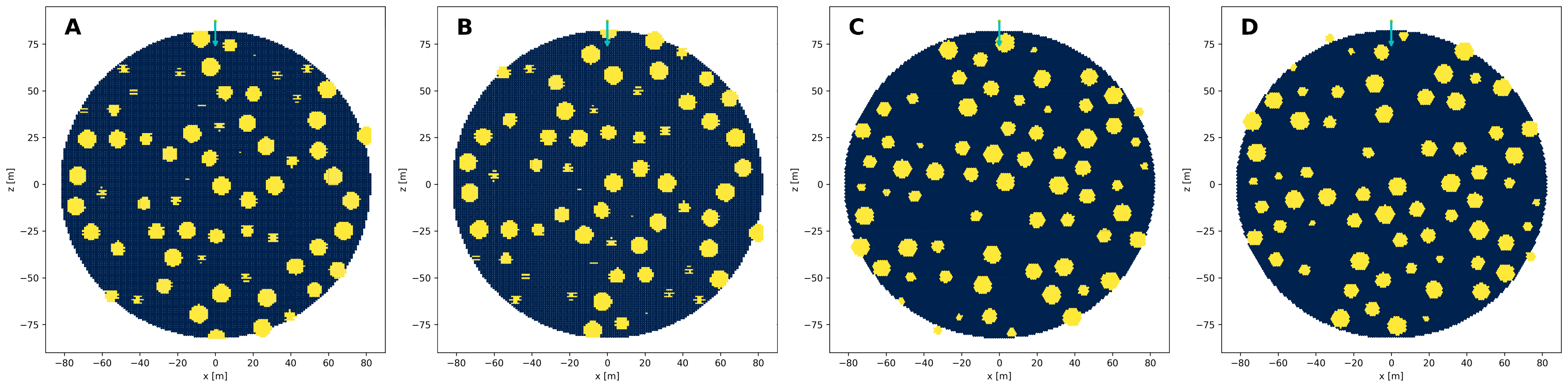}
        \caption{Slices through the initial target geometries for the loose boulder packing scenarios. Top row: 2.5 to 12 m boulders. Bottom row: 5 m boulders. The arrow at the top of each panel shows the impact location. } \label{fig:init2}
\end{figure*}

Here we provide slices through the initial target set-ups used in this study. Figures~\ref{fig:init_800}, ~\ref{fig:init_ryu}, and ~\ref{fig:init_300} show the initial target geometries for the high boulder packing scenarios, and Fig.~\ref{fig:init2} shows the initial target geometries for the loose boulder packing scenarios.

\section{Summary of simulations and results}

Here we provide a summary of our the initial conditions and results of our simulations (Table~\ref{table:results}).

\begin{table*}[h]
    \footnotesize
        \caption{Simulation summary. }
        \begin{tabular}{lccccccc}
    Description & Mass boulders & Mass matrix & Mass target & Boulder mass fraction & Packing & Beta & Outcome  \\
               & (kg) & (kg) & (kg) & (\%) & & & \\
    \hline
    Homogeneous & 0 & 3.99$\times$10$^9$ & 3.99$\times$10$^9$ & 0.00 & -- & 4.34 & Global def.\\
    \hline 
    2.5 m A & \multirow{4}{*}{1.47$\times$10$^9$} & \multirow{4}{*}{2.65$\times$10$^9$} & \multirow{4}{*}{4.12$\times$10$^9$} & \multirow{4}{*}{35.70} & \multirow{4}{*}{high}& 1.62 & Cratering\\
    2.5 m B &  & & & & & 1.64 & Cratering \\
    2.5 m C &  & & & & & 1.90 & Cratering \\
    2.5 m D &  & & & & & 2.14 & Cratering \\
    \hline 
    5 m A & \multirow{4}{*}{6.55$\times$10$^8$} & \multirow{4}{*}{3.40$\times$10$^9$} & \multirow{4}{*}{4.05$\times$10$^9$} & \multirow{4}{*}{16.17} & \multirow{4}{*}{loose} & 4.35 & Global def.\\
    5 m B & &  & & & &  4.72 & Global def.\\
    5 m C & &  & & & &  2.98 & Global def.\\
    5 m D & &  & & & &  5.13 & Global def.\\
    \hline
    8 m A & \multirow{4}{*}{2.21$\times$10$^9$} & \multirow{4}{*}{2.00$\times$10$^9$} & \multirow{4}{*}{4.21$\times$10$^9$} & \multirow{4}{*}{52.36} & \multirow{4}{*}{high} & 2.07 & Cratering \\
    8 m B & &  & & & &  1.51 & Cratering \\
    8 m C & &  & & & &  1.89 & Cratering \\
    8 m D & &  & & & &  2.64 & Cratering \\
    \hline
    2.5 -- 14 m A & \multirow{4}{*}{1.54$\times$10$^9$} & \multirow{4}{*}{2.50$\times$10$^9$} & \multirow{4}{*}{4.04$\times$10$^9$} & \multirow{4}{*}{38.18} & \multirow{4}{*}{high} & 2.86 & Cratering \\
    2.5 -- 14 m B & &  & & & &  1.90 & Cratering \\
    2.5 -- 14 m C & &  & & & &  3.15 & Cratering \\
    2.5 -- 14 m D & &  & & & &  2.48 & Cratering \\
    \hline 
    2.5 -- 12 m A & \multirow{4}{*}{7.01$\times$10$^8$} & \multirow{4}{*}{3.31$\times$10$^9$} & \multirow{4}{*}{4.01$\times$10$^9$} & \multirow{4}{*}{17.50} & \multirow{4}{*}{loose} & 4.30 & Global def. \\
    2.5 -- 12 m B & &  & & & &  4.45 & Global def.\\
    2.5 -- 12 m C & &  & & & &  5.22 & Global def.\\
    2.5 -- 12 m D & &  & & & &  4.15 & Global def.\\
        \hline
        \end{tabular}
        \label{table:results}
\end{table*}

\end{document}